\documentclass[sigconf]{acmart}
\usepackage[utf8]{inputenc}
\usepackage{fancyvrb}
\usepackage[store-all]{keytheorems}

  \InputIfFileExists{no-editing-marks}{
    \def\noeditingmarks{}
  }

  \usepackage{xargs}
  \usepackage[colorinlistoftodos,prependcaption,textsize=tiny]{todonotes}
  \ifx\noeditingmarks\undefined
      \newcommandx{\unsure}[2][1=]{\todo[linecolor=orange,backgroundcolor=orange!25,bordercolor=orange,#1]{#2}}
      \newcommandx{\unsureF}[2][1=]{\todo[linecolor=orange,backgroundcolor=lightgray!25,bordercolor=orange,#1]{#2}}
      \newcommandx{\info}[2][1=]{\todo[linecolor=green,backgroundcolor=green!25,bordercolor=green,#1]{#2}}
      \newcommandx{\change}[2][1=]{\todo[linecolor=blue,backgroundcolor=blue!25,bordercolor=blue,#1]{#2}}
      \newcommandx{\inconsistent}[2][1=]{\todo[linecolor=red,backgroundcolor=red!25,bordercolor=red,#1]{#2}}
      \newcommandx{\critical}[2][1=]{\todo[linecolor=purple,backgroundcolor=purple!25,bordercolor=purple,#1]{#2}}
      \newcommand{\improvement}[1]{\todo[linecolor=pink,backgroundcolor=pink!25,bordercolor=pink]{#1}}
      \newcommandx{\resolved}[2][1=]{\todo[linecolor=OliveGreen,backgroundcolor=OliveGreen!25,bordercolor=OliveGreen,#1]{#2}} 
    \else
      \renewcommand{\todo}{}
      \newcommandx{\unsure}[2][1=]{{}}
      \newcommandx{\unsureF}[2][1=]{{}}
      \newcommandx{\info}[2][1=]{{}}
      \newcommandx{\change}[2][1=]{{}}
      \newcommandx{\inconsistent}[2][1=]{{}}
      \newcommandx{\critical}[2]{{}}
      \newcommand{\improvement}[1]{{}}
      \newcommandx{\resolved}[2][1=]{{}}

  \fi

\title{Refinement-Types Driven Development: A study}

\author{Facundo Domínguez}
\affiliation{
     \institution{Tweag}
     \country{France}
}
\email{facundo.dominguez@tweag.io}
\author{Arnaud Spiwack}
\affiliation{
     \institution{Tweag}
     \country{France}
}
\email{arnaud.spiwack@tweag.io}


\acmConference[IFL 2025]{37th Symposium on Implementation and Application of Functional Languages}{October 2025}{Montevideo, Uruguay}

\newcommand{\tc}[1]{{\small\texttt{#1}}}
\newcommand{\codeblocksize}{\fontsize{6.5}{9}\selectfont}
\newcommand{\sourcefile}[1]{\tc{#1}\footnote{\scriptsize\url{https://github.com/tweag/ifl2025-liquidhaskell/blob/main/src/examples/#1}}}
\newcommand{\patchfile}[1]{\tc{#1}\footnote{\scriptsize\url{https://github.com/tweag/ifl2025-liquidhaskell/blob/main/src/patches/#1}}}
\RecustomVerbatimEnvironment{verbatim}{Verbatim}{
    fontsize=\codeblocksize,
}
\newtheorem{principle}{Principle}

\begin{document}
\begin{abstract}
    This paper advocates for the broader application of SMT solvers in everyday
    programming, challenging the conventional wisdom that these tools are
    solely for formal methods and verification. We claim that SMT solvers, when
    seamlessly integrated into a compiler's static checks, significantly
    enhance the capabilities of ordinary type checkers in program composition.
    Specifically, we argue that refinement types, as embodied by Liquid Haskell,
    enable the use of SMT solvers in mundane programming tasks.

    Through a case study on handling binder scopes in compilers, we envision a
    future where ordinary programming is made simpler and more enjoyable with the
    aid of refinement types and SMT solvers. As a secondary contribution, we
    present a prototype implementation of a theory of finite maps for Liquid
    Haskell's solver, developed to support our case study.
\end{abstract}
\maketitle

\section{Introduction}

SMT solvers are useful to the ordinary activity of
programming. This is what we would like to convince the reader of. More
precisely, our claim is that an SMT solver, well-integrated in
a compiler, complements an ordinary type checker and can, in fact, be used much in
the same way. SMT solvers and type checkers are good at enforcing different kinds of
properties, broadening the ways in which we can design our programs.

SMT solvers, when it comes to their application to programming, are usually
paired in the literature with terms like ``formal methods'' or
``verification''~\cite{barnett05,demoura08,zinzin17,swamy22,filli13,leino17}. We
would like to challenge the wisdom that we reach for SMT-solver-based tools when
we need formal methods. We would benefit from using SMT solvers in mundane
programs. Not because it makes programs more correct, but because it helps us
write the programs we want.

We will be arguing, in particular, that refinement types, in the guise of Liquid
Haskell~\cite{vazou14b}, let you do just that. Even though Liquid
Haskell is also usually invoked together with phrases like ``formal methods'' or
``verification''~\cite{vazou14,lehmann21,liu20,redmond23}.

Through a case study, we will argue for a future where programming, ordinary
programming, is made easier and more pleasant thanks to refinement types and
SMT solvers, even though the technology is not ready yet, as we discuss
in Section~\ref{evaluation}. Our case study
will be the handling of binders' scopes in compilers. We distill from the experience
a set of principles that were useful to us and which could apply to other scenarios with
this programming style.
A secondary contribution
is a prototype implementation of a theory of finite maps for Liquid Haskell's
solver, to support our case study, and which we discuss in
Section~\ref{extending-liquid-haskell}.

\section{Capture-avoiding substitutions}
\label{capture-avoiding-substitution}

Binding scope management is recognized as a persistent annoyance when writing compilers.
It is easy to get wrong and it is a source of mistakes to the point that many have
proposed disciplines to prevent mismanagement of scopes.
The canonical mistake example is name capture in substitutions like
$(\lambda x. y)[y:=t]$. The result of this substitution is $\lambda x. t$.
Thus $(\lambda x. y)[y:=x]$ is $\lambda x. x$. An easy mistake!

Compiler authors have proposed many disciplines to help make scope more
manageable.
The GHC Haskell compiler, for instance, uses an approach to avoid name capture called
\textit{the rapier}~\cite{peytonjones02secrets}. All term-manipulating functions
carry an additional \textit{scope} set containing all the
variables that appear free in its arguments. This set is
used both to decide what to rename a binder to, in order to avoid name capture,
and it is also used to skip renaming a binder if it would not capture any free
variables. Figure~\ref{rapier-style-substitution} shows an implementation of
substitution
for the untyped lambda calculus.

\begin{figure}
\begin{verbatim}
data Exp = Var Int | App Exp Exp | Lam Int Exp

substitute :: Set Int -> Subst Exp -> Exp -> Exp
substitute scope s e0 = case e0 of
  Var i -> lookupSubst s i
  App e0 e1 -> App (substitute scope s e0) (substitute scope s e1)
  Lam i e
    | member i scope,
      let j = freshVar scope ->
        Lam j $ substitute (insert j scope) (extendSubst s i (Var j)) e
    | otherwise ->
        Lam i $ substitute (insert i scope) (extendSubst s i (Var i)) e

freshVar :: Set Int -> Int
freshVar s = case lookupMax s of Nothing -> 0; Just i -> i + 1
\end{verbatim}
\caption{Rapier style substitution}
\Description{Haskell implementation of substitution for untyped lambda terms using the rapier}
\label{rapier-style-substitution}
\end{figure}

\subsection{The foil}
\label{the-rapier-with-stronger-types}

The rapier was not enough, however, for \citet{maclaurin23} who report that
despite using the rapier they struggled with frequent scope issues in their
compiler. They set out to enforce the scope properties of the rapier with
Haskell's type system. A stunt that has often been attempted, but
\citeauthor{maclaurin23}'s approach, that they name \emph{the foil}, is probably the
first to succeed at enforcing such invariants without incurring an unreasonable
amount of boilerplate.
In Section~\ref{the-rapier-with-refinement-types}, we will argue that we can
achieve similar guarantees more economically with SMT solvers.

Here is our distillation of the properties that \citeauthor{maclaurin23} set
out to guarantee (see also \cite[Section~4]{maclaurin23}):
\begin{enumerate}
\item Every traversed binder must be added to the scope set, otherwise its name
      could be accidentally used later where a fresh name was intended.
\item \label{req:always-rename} Every traversed binder must be renamed if it is already a member of the
      scope set, because this name could otherwise be captured as above.
\item When renaming a binder, the new name must not belong to the scope set.
\item When renaming a binder, the occurrences of the old bound variable need
      to be substituted with the new name.
\item The initial scope set must contain the free variables in the input term
      and in the range of the substitution to apply.
\end{enumerate}

These properties are exigent, though they do not ensure that we can only write correct
substitution functions. For instance, with all these properties it's possible
to write a function which takes $(x\ y)[x:=x]$ to $(y\ x)$.
But as anticipated in the
introduction, we are not concerned with full correctness.

\Citeauthor{maclaurin23} propose a library with types \tc{Scope n}, \tc{Name n}, and
\tc{Name\-Binder n l}. A value of type \tc{Scope n} is a set of names, where
the type index \tc{n} is the name of the set at the type level. A value of type \tc{Name n} is a name that
belongs to the scope set \tc{n}. A value of type \tc{NameBinder n l} is
a name \tc{b} such that adding \tc{b} to scope set \tc{n} results in the scope
set \tc{l}.
These types are to be used in
the abstract syntax tree of terms:

\begin{quotation}
\begin{verbatim}
data Exp n = Var (Name n)
           | App (Exp n) (Exp n)
           | forall l. Lam (NameBinder n l) (Exp l)
\end{verbatim}
\end{quotation}

Then the operations and type checking on the new types will guide the user into
respecting the scope requirements when implementing substitution.

\begin{verbatim}
substitute :: Distinct o => Scope o -> Subst Expr i o -> Expr i -> Expr o
\end{verbatim}

This type signature says that no names shadow each other in the scope set \tc{o}.
It also says that the substitution will take an expression with free variables in
a scope set \tc{i} and produce an expression with free variables in a scope set
\tc{o}.

There
are mechanisms to check that a scope set is a subset of another, to assert that no
name shadows another one in a given scope set, to reason that expressions
with free variables in one scope (\tc{Exp n}) can be coerced to expressions with
free variables in a superset (\tc{Exp l}), and to introduce scope sets that extend
others with freshly created names. They also provide an implementation of maps of
variables to expressions, that is the substitutions to apply, with an interface
that uses the new types as well. There is for instance the following function to
produce fresh variables:

\begin{verbatim}
withRefreshed
  :: Distinct o
  => Scope o
  -> Name i
  -> (forall (o' :: S). DExt o o' => NameBinder o o' -> r)
  -> r
\end{verbatim}

Using the constraint \tc{DExt}, this type signature says that scope set \tc{o'}
extends the scope set \tc{o} with the given \tc{NameBinder o o'}. This binder
may have the same name as the provided \tc{Name i} if it was not present in
\tc{o}, otherwise it will be a fresh name. As another example, the following
function always produces a fresh name.

\begin{verbatim}
withFresh
  :: Distinct n
  => Scope n
  -> (forall l . DExt n l => NameBinder n l -> r )
  -> r
\end{verbatim}

With ingenious engineering and design, the foil meets its rather ambitious goal.
But it is unfortunate that the authors needed to be ingenious. All things equal,
we prefer program components to be straightforward. Because ingenious solutions
take time, and because straightforward solutions are easier to adapt when the
parameters of the problem evolve.

\subsection{A Liquid Haskell primer}

We will turn next to Liquid Haskell as our proposed solution, but first let us
introduce Liquid Haskell briefly.
Liquid Haskell is a
plugin for Haskell which statically checks that programs respect signatures
provided by the programmer. There are two key differences between Liquid Haskell
signature checking and a classical type checker:

\begin{itemize}
  \item The checking process consists in generating logical constraints or proof
        obligations which are then fed to an SMT solver, leveraging the powerful
        capabilities of SMT solvers to reason about numbers, arrays, strings,
        and other sorts.
  \item Signatures are expressed with \emph{refinement types} of the form
        \tc{\{x:b | p\}}, which denote values of base type \tc{b} that
        satisfy predicate \tc{p}. We will write sometimes \tc{b<p>} to denote
        \tc{\{x:b | p x\}}.
Refinements are subject to subtyping in the same way as subsets in set
theory, so that we have
\begin{verbatim}
{-@ f :: {x:Int | x > 1} -> {x:Int | x > 0} @-}
f :: Int -> Int
f x = x
\end{verbatim}
\end{itemize}

Liquid Haskell reads refinement type signatures and other annotations from
inside special Haskell comments
\tc{\{-@ \ldots\ @-\}}. We will skip them in our snippets when it is unambiguous.

The predicates in the refinement types are in a language of expressions
referred to as the logic language. For the sake of this paper, we can
regard it as a subset of Haskell, except that predicates are assembled both from
regular Haskell functions and functions that are
only available in the logic language.

We will use sparingly the following form of refinement type signature.
\begin{verbatim}
{-@ idInt :: forall <p :: Int -> Bool>. Int<p> -> Int<p> @-}
idInt :: Int -> Int
idInt x = x
\end{verbatim}
We say that \tc{p} is an abstract predicate, and it is inferred by Liquid Haskell
depending on the context in which \tc{idInt} is used.

A function like \tc{member}, which comes from the module \tc{Data.Set}
in the \tc{containers} package, is linked by Liquid Haskell to the
SMT solver's theory of sets.
\begin{verbatim}
import Data.Set
assume member :: Ord a
              => x:a -> xs:(Set a) -> {v:Bool | v <=> Set_mem x xs}
\end{verbatim}
Refinement type signatures starting with the \tc{assume} keyword declare that the
corresponding Haskell function honors the signature, but it is not
checked. In this case, it is because \tc{Data.Set} is an external dependency that
Liquid Haskell can not check. But it can also be applied to our own functions.

Here \tc{Set\_mem} is a symbol that Liquid Haskell maps to the theory
of sets in the SMT solver. While Liquid Haskell does not check that
\tc{member} behaves as declared in the refinement type signature,
it will assume the property in the return refinement type
whenever \tc{member} is used in a program.

Notice how the predicate on the return type mentions both arguments. Liquid
Haskell lets us express refinement types which relate arguments with each other,
and with the result in this manner. This obviates the need to
give a type-level name to arguments using existential quantification.

To define a function only available to use in Liquid Haskell annotations,
we can use the \tc{measure} keyword, such as:
\begin{verbatim}
measure listElts :: [a] -> Set a
  listElts []     = {v | (Set_emp v)}
  listElts (x:xs) = {v | v = Set_cup (Set_sng x) (listElts xs) }
\end{verbatim}
Here \tc{Set\_cup} and \tc{Set\_sng} are predefined functions to express the
union of sets and the singleton set respectively.

It is also possible to define uninterpreted symbols by simply omitting the
definition. It would look like this
\begin{verbatim}
measure listElts :: [a] -> Set a
\end{verbatim}
The meaning of the function would then be given by \tc{assume} refinement
type signatures on other functions. See for instance the use of the \tc{domain}
function in the following section.

\subsection{The rapier, refined}
\label{the-rapier-with-refinement-types}

We argue, next, that using Liquid Haskell to enforce the
requirements from Section~\ref{the-rapier-with-stronger-types} is more
straightforward than using the type checker alone. The code presented in this section is available in the file
\sourcefile{Subst1.hs}.

We define a function \tc{freeVars} in the same module as \tc{subs\-ti\-tute},
which collects the free variables of an expression. We note that this function
is only used in refinement type signatures, and in particular, it is not evaluated
when calling to \tc{substitute}.

\begin{verbatim}
freeVars :: Exp -> Set Int
freeVars e = case e of
  Var i -> singleton i
  App e1 e2 -> union (freeVars e1) (freeVars e2)
  Lam i e -> difference (freeVars e) (singleton i)
\end{verbatim}

Next, we need to give the following refined signature to the \tc{freshVar} of
Figure~\ref{rapier-style-substitution}:
\begin{verbatim}
{-@ assume freshVar :: s:Set Int -> {v:Int | not (member v s)} @-}
\end{verbatim}
This signature is assumed rather than checked. We could choose to check it, but
Liquid Haskell does not have a good built-in understanding of the \tc{lookupMax}
function that we use. So instead, we choose to assume the signature. This is our
first principle of programming with refinement types:

\begin{principle}
\label{assumption-principle}
  Typically, refinement types allow you to reduce the trusted code base, but they also offer
    you a choice. When it is easier to prove a result by
    hand than with the SMT solver, you can assume the property and
    justify it informally.
\end{principle}

In this article, by \emph{trusted code base}, we mean the portion of a
codebase where the programmer must prove the desired properties
herself rather than relying on static checks to enforce said properties.
Tooling like compilers, type checkers, SMT solvers, and operative systems
are excluded from this definition.

It is good discipline to justify systematically why assumptions should hold.
An incorrect assumption could make Liquid Haskell accept programs that do not
meet the properties we mean to check. The consequences range through the whole
gamut from incorrect results, to security vulnerabilities and crashes, depending
on the kind of checks.

Finally, we will take as a parameter a datatype representing substitutions
(\emph{i.e.} finite maps of variables to terms). To represent this parameter in
our study we take an abstract type and \tc{assume} the necessary properties that a
substitution type needs to respect. Since this is ordinary programming, not a
verification project, we need to test our code, and we provide a concrete type
for that sake. But using an abstract type ensures that we can support
any efficient substitution type.

\begin{verbatim}
data Subst t -- opaque
{-@ measure domain :: Subst e -> Set Int @-}

assume lookupSubst
  :: forall <p :: Exp -> Bool>.
     s:Subst Exp<p>
  -> {k:Int | member k (domain s)}
  -> Exp<p>

assume extendSubst
  :: s:Subst a
  -> i:Int
  -> a
  -> {v:Subst a | union (domain s) (singleton i) = domain v }
\end{verbatim}

Notice that the logical function \tc{domain}, which stands for the set of
variables that the substitution defines, is uninterpreted. It must be since it is an
assumption.

That's it, this is the entirety of our trusted code base for this example. For
the most part, it required thinking about what properties we wanted to enforce,
but not much about how they ought to be enforced.

In order to deal with scope checks, we define a type alias \tc{ScopeExp S},
that is the type of all
expressions whose free variables are in the set \tc{S}\footnote{In type aliases,
Liquid Haskell expects parameter names corresponding to terms (i.e. not types) to start with
an uppercase letter.}.

\begin{verbatim}
{-@ type ScopedExp S = {e:Exp | isSubsetOf (freeVars e) S} @-}
\end{verbatim}
Functions like \tc{isSubsetOf} and \tc{difference} come from the \tc{Data.\allowbreak Set}
module. We can give now the following signature to \tc{substitute}
\begin{verbatim}
{-@
substitute
  :: scope:Set Int
  -> s:Subst (ScopedExp scope)
  -> ScopedExp (domain s)
  -> ScopedExp scope
@-}
substitute :: Set Int -> Subst Exp -> Exp -> Exp
\end{verbatim}
Remarkably, this implementation for \tc{substitute}, where we check static scopes,
is unchanged from the implementation of
Figure~\ref{rapier-style-substitution}. This will not always be the case, but this
exemplifies how using Liquid Haskell to enforce invariants tends to create less
boilerplate than a type-based approach.

The refinement type signature of \tc{substitute} is a direct translation of the
Haskell type signature used by the foil.
\begin{verbatim}
substitute :: Distinct o => Scope o -> Subst Expr i o -> Expr i -> Expr o
\end{verbatim}
The foil's \tc{Scope o} type becomes a
regular set \tc{scope:Set Int} of names, there's no need for the type parameter
\tc{o}, which the foil uses as a type-level name for the scope, since we can directly refer to \tc{scope} in terms. The foil's \tc{Subst Expr i o} type
becomes \tc{s:Subst (ScopedExp scope)}, the parameter \tc{i} is omitted and
referred to as \tc{domain s} instead. The foil's \tc{Expr i} type becomes
\tc{ScopedExp (domain s)}, which still requires the free variables of the input
expression to be in the domain of the substitution. And finally, both return types
\tc{Expr o} and \tc{ScopedExp scope} require the free variables of the output to
be in the given scope set.

Figure~\ref{rapier-style-substitution} uses that a substitution
\tc{s :: Subst (ScopedExp scope)}
also has (refined) type
\tc{s :: Subst (ScopedExp (insert i scope))},
as there are recursive calls like
\begin{verbatim}
substitute (insert i scope) (extendSubst s i (Var i))
\end{verbatim}
which requires
\begin{verbatim}
extendSubst s i (Var i) :: Subst (ScopedExp (insert i scope))
\end{verbatim}
which in turn requires
\begin{verbatim}
s :: Subst (ScopedExp (insert i scope))
\end{verbatim}
This kind of subtyping is trivial with refinement types. It is the default
behavior. Whereas with an ML type system, subtyping is not a typical feature. The
foil, for instance, needs an explicit function to cast substitutions when
extending a scope. This is our next principle:
\begin{principle}
  Refinement types add a layer of subtyping on top of your type system. When
  your program is best modeled with subtyping you should consider refinement
  types.
\end{principle}

The type of lambda terms is also unchanged, as the well-scoping invariant is
applied to a whole term at once. A nice consequence of it is that functions
that do not benefit from all the scope checking business can simply take a naked
term and ignore it. The \tc{freeVars} function, for example, is implemented on
naked terms.

\subsection{A hybrid approach}
\label{ensuring-the-scope-set-is-checked}

Our refinement type signature of \tc{substitute} follows the type signature of
\citeauthor{maclaurin23} to the letter.
Yet we can introduce the following bug in \tc{substitute} from
Figure~\ref{rapier-style-substitution}, where we omit the fresh binder \tc{j}:
\begin{verbatim}
  ...
  Lam i e
    | member i scope ->
      Lam i $ substitute (insert i scope) (extendSubst s i (Var i)) e
    | otherwise -> ...
\end{verbatim}
Liquid Haskell flags no errors but the program will still misbehave as
follows (in pseudo-Haskell).

$$\codeblocksize{\tc{substitute}~\{\tc{x}\}~(\lambda{}\tc{x}. \tc{y}) [\tc{y}:=\tc{x}] = (\lambda{}\tc{x}. \tc{x})}$$

What is going on? The binder \tc{i} is now capturing free variables in the
range of the substitution. The signature is, in fact, indifferent to whether
the binder \tc{i} is already present or not in the scope set. There is no
mechanism to prevent adding a binder that is already present in the scope set.
That is, we fail to enforce Property~(\ref{req:always-rename}) from Section~\ref{the-rapier-with-stronger-types}.
And, more to the point, how could we? “Never add a binder to the scope set that is already
present” is not a set theoretical property. It is not even a functional property.
It is a kind of temporal invariant.

Such temporal invariants are not naturally expressed in the logic of Liquid Haskell.
But they are quite easy to implement with abstract types. So let us use an abstract
type. What we need to do is to ensure that whenever we see a new binder it must
be tested against the scope, and that this test is packaged together with fresh
name generation.

We follow the foil and
introduce an abstract type \tc{Scope} and a function \tc{withRefreshed}. The types are a little
simpler because we do not need existential quantification to reflect value-level
objects at the type level, but otherwise these are the same functions and types
as in Section~\ref{the-rapier-with-stronger-types}.
\begin{verbatim}
newtype Scope = UnsafeScope { unsafeUnScope :: (Set Int) }
{-@
predicate Member E S = Set.member E (unsafeUnScope S)

withRefreshed :: s:Scope -> i:Int
  -> {p:(Scope, Int) |
       not (Member (snd p) s) && fst p == union s (singleton (snd p))}
@-}
withRefreshed :: Scope -> Int -> (Scope, Int)
withRefreshed (UnsafeScope s) i
  | Set.member i s = let j = freshVar s in (UnsafeScope (insert j s), j)
  | otherwise = (UnsafeScope (insert i s), i)
\end{verbatim}

We needed to add a refinement type signature to \tc{withRefreshed} to serve as
glue with the Liquid Haskell world. This refinement type signature tells Liquid
Haskell precisely that \tc{withRefreshed} does both membership checking and
fresh variable call: the variable returned by \tc{withRefreshed} is not in the
old scope but is in the new scope.

We make the type \tc{Scope} abstract to enforce that binders are always
refreshed when traversed, as \tc{withRefreshed} is the only
way to test for membership and to extend a scope. This is why we define a
\tc{Member} predicate alias, only available in the logic, but provide no
\tc{member} function in Haskell for \tc{Scope}s.
The full code for this example can be found in the
file \sourcefile{Subst2.hs}.

This is our next principle for refinement types:
\begin{principle}
  Refinement types and abstract types are best at enforcing different kind of
  properties. You should use the simpler solution for each property that you
  need, as refinement types and abstract types mix well.
\end{principle}

\section{Unification}
\label{unification}

Now that we have established the refined rapier interface, let us show how it
can be applied to a more realistic example: solving first-order equational
formulas. Specifically, we will be solving a form of Horn clauses in
the Herbrand domain. This is the sort of unification problem which can show up
when type-checking programs with GADTs~\cite{schrijvers09}. Scope management in
such a solver is a much trickier business than in the case of mere substitutions
and, in the authors' experience, something where any help from the compiler is welcome.
The source code of this section can be found in the file
\sourcefile{Unif.hs}.

In addition to variables, still represented as integers, we have unification
variables. Unification variables have their own scopes: the formula
$\exists x. \forall y. x=y$ does not have a solution. It will be reduced to a
formula of the form $f_{x} = y$ where $f_{x}$ is a unification variable; we very much
don't want this unification problem to succeed: we shall make it so that $y$ is not in
the permissible scope for $f_{x}$.

Furthermore, the unification algorithm will perform substitutions. Substitutions are blocked by
unification variables as we do not know what they stand for yet. So a unification
variable, in our syntax, is a pair $(f, [x_0:=t_0,\ldots,x_n:=t_n])$ of a
unification variable proper and a suspended substitution. Where
$\{x_0,\ldots,x_{n}\}$ is the scope of $f$. Such a pair is akin to a skolem
function application $f(t_0,\ldots,t_n)$. Notice in particular, how the solution
of $f$ can only have free variables in $\{x_0,\ldots,x_{n}\}$, but
$(f, [x_0:=t_0,\ldots,x_n:=t_n])$ may live in a different scope altogether.
This type of unification problem is tricky because there are multiple
intermingled scopes to manage, rather than one like
in the case of substitution (Section~\ref{capture-avoiding-substitution}).

\begin{verbatim}
type Var = Int
type SkolemApp = (Var, Subst Term)
\end{verbatim}

This way, our formula $\exists x. \forall y. x=y$ will be reduced to
$(f_{x},[]) = y$ which does not have a solution. On the other hand
$\forall x. \exists y. x = y$ becomes $x = (f_{y}, [x:=x])$ so $x$ is a solution
for $f_{y}$ and the formula is solvable.

Our unification algorithm is a first-order variant of pattern
unification~\cite{miller91-pattern} sufficient to eliminate equalities to the
left of implication in the style proposed by~\citet{miller22}. The main
functions, sans refined signatures, can be found in
Figure~\ref{conditional-unification}. Unification algorithms can get pretty
finicky, for the sake of simplicity our algorithm is not as complete as it could
be and will miss some solutions\footnote{We have, on the other hand, tried to
  make the algorithm correct, so if it finds unsound solution it is a bug and we
  apologize.}.

At the heart of the algorithm is substitution inversion~\cite{ziliani15}: when
encountering an equality of the form
$$(f_{x},[y:=a, z:=b]) = u$$
If there is a solution, we want it to be
$$f_{x} := u[a:=y, b:=z]$$
This is the same as pattern unification, except that it does not need terms to
contain functions. The \tc{inverseSubst} function is responsible for this
inversion.

We are choosing a language of terms with both regular variables (representing
variables bound by universal quantifiers), skolem applications representing
unification variables with their substitutions, and sufficient
constructors to encode arbitrary terms. Here is the concrete type of terms, as
well as that of formulas where the only thing to remark is that the left-hand
side of implications is a single equality.

\begin{verbatim}
data Term
  = V Var | SA SkolemApp | U | L Term | P Term Term

data Formula
  = Eq Term Term               -- equality
  | Conj Formula Formula       -- conjunctions
  | Then (Term, Term) Formula  -- a = b => f
  | Exists Var Formula         -- existential quantification
  | Forall Var Formula         -- universal quantification
\end{verbatim}

In Figure~\ref{conditional-unification}, the function \tc{unify} takes a rapier scope parameter containing all the variables that
can appear free in the input formula. This set is used to rename \tc{Forall}
binders when doing substitutions. For instance, unifying the following formula
$$\forall x. \forall y. \exists z. y = L(x) \Rightarrow \forall x. y = z$$
reduces to unifying
$$\forall x. \forall y. \exists z. (\forall x. y = z)[y:=L(x)]$$
and the substitution needs to rename the inner binder $x$.

In a preceding pass (Section~\ref{sec:skolemize}), existential quantifiers are replaced
with skolem applications, so in \tc{unify} we assume that there is no
existential quantifier. We have functions \tc{substituteFormula} and \tc{substitute}
to apply substitutions in formulas and terms respectively, and \tc{substituteSkolems} to  substitute
unification variables in formulas. We have a function \tc{skolemSet} to collect the skolem applications of a
term. And a function \tc{fromListSubst} to construct a substitution from a list of
pairs \tc{[(Var, Term)]}.

The functions \tc{substEq} and \tc{unifyEq} are simplified here for the sake of presentation.
They handle more cases in the reference source code, but these cases are not essential to our
discussion.

\begin{figure}
\begin{verbatim}
unify :: Set Int -> Formula -> Maybe [(Var, Term)]
unify s (Forall v f) = unify (Set.insert v s) f
unify s (Exists v f) = error "unify: the formula has not been skolemized"
unify s (Conj f1 f2) = do
    unifyF1 <- unify s f1
    unifyF2 <- unify s (substituteSkolems f2 unifyF1)
    return (unifyF1 ++ unifyF2)
unify s f@(Then (t0, t1) f2) =
    let subst = fromListSubst (substEq t0 t1)
     in unify s (substituteFormula s subst f2)
unify s (Eq t0 t1) = unifyEq t0 t1

substEq :: Term -> Term -> [(Var, Term)]
substEq (V i) t1 = [(i, t1)]
substEq t0 (V i) = [(i, t0)]
substEq _ _ = []

unifyEq :: Term -> Term -> Maybe [(Var, Term)]
unifyEq t0 t1@(SA (i, s))
  | Just s' <- inverseSubst $ narrowForInvertibility (freeVars t0) s
  , let t' = substitute s' t0
  , not (Set.member i (skolemSet t'))
  , Set.isSubsetOf (freeVars t') (domain s)
  = Just [(i, t')]
unifyEq t0@(SA _) t1 = unifyEq t1 t0
unifyEq _ _ = Nothing

-- | @narrowForInvertibility vs s@ removes pairs from @s@ if the
-- range is not a variable, or if the range is not a member of @vs@.
narrowForInvertibility :: Set Var -> Subst Term -> Subst Term
narrowForInvertibility vs (Subst xs) =
  Subst [(i, V j) | (i, V j) <- xs, Set.member j vs]

inverseSubst :: Subst Term -> Maybe (Subst Term)
inverseSubst (Subst xs) = fmap Subst (go xs)
  where
    go [] = Just []
    go ((i, V j) : xs) = fmap ((j, V i) :) (go xs)
    go _ = Nothing
\end{verbatim}
\caption{Conditional unification}
\Description{Haskell implementation of condition unification}
\label{conditional-unification}
\end{figure}

The function \tc{unifyEq} defines what a good solution should be.
One of the conditions is that whatever term \tc{t'} is proposed
as solution for a skolem \tc{i}, it needs to have as free variables only those in the
domain of the substitution defining the skolem application
(\textit{scope check}). For instance, in $(f_x , [x:=y]) = P(y,y)$,
$P(x,x)$ is a solution that satisfies the scope check, but $P(x,y)$ would be a solution
that doesn't since $y$ is not in the domain of $[x:=y]$.

Another condition is that the skolem \tc{i} should not occur in the solution
\tc{t'} (\textit{occurs check}). For instance, in the previous example
$f_x:=P(x,f_x)$ is a solution that doesn't pass the occurs check. In addition,
since we are inverting a substitution to find
\tc{t'}, we might not find solutions if we cannot invert the
substitution. This implementation only inverts substitutions where
variables are mapped to variables. That is, we solve $(f, [z:=x]) = L(L(x))$
to get the solution $f:=L(L(z))$ but we do not try solving, say, $(f, [z:=L(x)]) = L(L(x)))$.

\subsection{A look at skolemization}
\label{sec:skolemize}

Figure~\ref{skolemization} shows the function to replace existential quantifiers
with unification variables. This example is interesting because the complexity of
managing the scopes for both universal and existential quantifiers
considerably exceeds the canonical example of the rapier.

\begin{figure}
\begin{verbatim}
skolemize :: Set Int -> Formula -> State (IntMap (Set Int)) Formula
skolemize sf (Forall v f) = do
    m <- get
    put (IntMap.insert v sf m)
    f' <- skolemize (Set.insert v sf) f
    pure (Forall v f')
skolemize sf (Exists v f) = do
    m <- get
    let u = if IntMap.member v m then
              freshVar (Set.fromList (IntMap.keys m))
            else
              v
        m' = IntMap.insert u sf m
    put m'
    let subst = fromListSubst [(v, SA (u, fromSetIdSubst sf))]
    skolemize sf (substituteFormula sf m' subst f)
skolemize sf (Conj f1 f2) = do
     f1' <- skolemize sf f1
     f2' <- skolemize sf f2
     pure (Conj f1' f2')
skolemize sf f@(Then (t0, t1) f2) = do
     f2' <- skolemize sf f2
     pure (Then (t0, t1) f2')
skolemize _ f@Eq{} = pure f
\end{verbatim}
\caption{Skolemization}
\Description{Haskell implementation of skolemization}
\label{skolemization}
\end{figure}

The \tc{skolemize} function takes a set \tc{sf} as an argument as well as a
finite map \tc{m} as the state of a state monad. The set \tc{sf} is the scope set
of variables that have been
introduced with universal quantification, and can appear free in the
input formula. The finite map \tc{m} contains the variables
that have been introduced with existential quantification together with their own
scopes, that is, the universally quantified variables in scope at the original existential
binder.

We pass the map \tc{m} as a monadic state, because we do not
want to generate the same unification variable for existential binders appearing
on different subformulas, since unification variables scope over the entire formula. For instance, the following formula
$$\forall x. \exists y. x = y \land \forall z. \exists y. z = y$$
should produce unification variables like
$$\forall x. x = y[x:=x] \land \forall z. z = w[x:=x, z:=z]$$
It would be a mistake to call both unification variables $y$ and $w$ the same.
Their occurrences even have different scopes!

We expect the set \tc{sf} to be a subset of the keys in \tc{m}. This is to reflect the
fact that, for debugging purposes, we do not want unification variables to be called the same as
universally quantified variables. It is not a strict requirement, but one
that makes the output of \tc{skolemize} considerably easier to read.

Yet, we do need to keep the scope set \tc{sf} separate from the monadic
state because it is needed to construct the skolem function applications where
existential variables are found.

Here is the refinement type signature of \tc{skolemize}.
\begin{verbatim}
type ScopedFormula S = {f:Formula | isSubsetOf (freeVarsFormula f) S}

assume skolemize
  :: sf:Set Int
  -> f:ScopedFormula sf
  -> State
       <{\m0 ->
            isSubsetOf sf (IntMapSetInt_keys m0)
         && consistentScopes m0 f
        }

       , {\m0 v m ->
             consistentScopes m v
          && existsCount v = 0
          && isSubsetOf (freeVarsFormula v) sf
          && intMapIsSubsetOf m0 m
         }>
       (IntMap (Set Int)) Formula
\end{verbatim}

This type signature is, admittedly, a bit involved. However while we were
designing this case study, \tc{skolemize} stayed without a refined signature
until pretty much the very end. This is possible because the inherent subtyping
of refinement types makes it easy to use unrefined and refined functions together.
Of course this prevented us from having guarantees for the program end-to-end,
but it is fine to add guarantees only where you need them. What you choose
to harden will not have to infect the rest of the program.
Which leads us to our next principle

\begin{principle}
  Functions with refined signature and without mix well. You should first use
  refinement types on function with the best power-to-weight ratio. You can
  incrementally add stronger types on more functions as your program evolves.
\end{principle}

Liquid Haskell helpfully lets us treat the state monad as equipped with
a Hoare logic \tc{State<pre,post>}. The supporting code for the refined state
monad is not readily available in Liquid Haskell. It probably should be, but in
the meantime, it can be found in Liquid Haskell's test suite, so we simply
copied it in the file \sourcefile{State.hs}.

The main conjuncts of the postconditon are \tc{consistent\-Scopes m v} and
\tc{existsCount v = 0}, the rest are invariants used by the recursive calls of
\tc{skolemize}.
\begin{itemize}
  \item \tc{existsCount v = 0} means that \tc{skolemize} returns a formula without
  existential quantifiers. As it is a requirement of \tc{unify}.
  \item \tc{consistentScopes m v} means that \tc{skolemize} returns a
  formula $F$ such that all the occurrences of any unification variable $i$ in $F$
  have an attached substitution whose domain is the scope of $i$ as reported by \tc{m}. This is our main scope invariant for this
  section.
\end{itemize}

While it is possible to define \tc{skolemize} with a set of unification variables
in the state instead of a finite map, the map choice makes easier to express the
consistency of the unification scopes. Changing the functions to make them easier
to explain is a topic which we will find again later on.

This signature for \tc{skolemize} cannot be checked with Liquid Haskell today
due to
a bug, so we ended up assuming the refinement type signature
in keeping with Principle~\ref{assumption-principle}. The rest
of the code does not benefit less because of it.

\subsection{The theory of \tc{unifyEq}}
\label{checking-unifyEq}

Let us now turn to the \tc{unifyEq} function, which is a traditional unification
function: it takes an equation and returns definitions for its unification
variables.
The refined signature that we give to \tc{unifyEq} statically enfoces scope checks, occurs checks,
and the consistency of scopes in the result and in the arguments.

\begin{verbatim}
type ConsistentScopedTerm S M =
  {t:Term | isSubsetOf (freeVars t) S && consistentScopesTerm M t}

unifyEq
  :: s:Set Int
  -> m:IntMap (Set Int)
  -> t0:ConsistentScopedTerm s m
  -> t1:ConsistentScopedTerm s m
  -> Maybe
      [( v :: Var
       , {t:Term |
             consistentScopesTerm m t}
          && isSubsetOfJust (freeVars t) (IntMap.lookup v m)
          && not (Set.member v (skolemSet t))
         }]
\end{verbatim}

The predicate \tc{consistentScopesTerm m t} is only used in refinement types, and
checks that the domains of the unification variables' substitutions in a term
\tc{t} are the scopes given by \tc{m}.

\begin{verbatim}
consistentScopesTerm :: IntMap (Set Int) -> Term -> Bool
consistentScopesTerm m (V _) = True
consistentScopesTerm m (SA (i, s)) =
       IntMap.lookup i m == Just (domain s)
    && consistentScopesSubst m s
consistentScopesTerm m U = True
consistentScopesTerm m (L t) = consistentScopesTerm m t
consistentScopesTerm m (P t0 t1) =
    consistentScopesTerm m t0 && consistentScopesTerm m t1

consistentScopesSubst :: IntMap (Set Int) -> Subst Term -> Bool
consistentScopesSubst m (Subst xs) =
    all (\(_, t) -> consistentScopesTerm m t) xs
\end{verbatim}

We would like to draw the reader's attention to the parameters of \tc{s}
and \tc{m} in the refinement type signature of the \tc{unifyEq} function,
conspicuously absent in the implementation of
Figure~\ref{conditional-unification}. This is because, in the source code,
we have extended the implementation of \tc{unifyEq} and many other functions
with these parameters. We could reconstruct these scope assumptions in the
functions' preconditions, but it is more involved, and requires a great deal more
lemmas to convince the SMT solver.

\begin{principle}
  It is easier to express properties and to use an SMT solver when assumptions
  are explicit rather than reconstructing assumptions that are implicit.
  Do not hesitate to pass assumptions as arguments to functions, even if those
  arguments are not used by the function.
\end{principle}

Note that compilers typically remove such obviously unused arguments during
compilation. GHC certainly does. So there is essentially no computational cost to
these extra arguments anyway.

\subsection{Totality and \tc{unify}}
\label{checking-unify}

There is not much more to add for the \tc{unify} function, but let us take this
opportunity to talk about the totality requirement. Here is its signature.

\begin{verbatim}
unify
  :: s:Set Int
  -> m:IntMap (Set Int)
  -> {f:ConsistentScopedFormula s m | existsCount f = 0}
  -> Maybe
       [( v :: Var
        , { t:Term |
              consistentScopesTerm m t
           && isSubsetOfJust (freeVars t) (IntMap.lookup v m)
           && not (Set.member v (skolemSet t))
          }
        )] / [formulaSize f]
\end{verbatim}

Notice the precondition \tc{existsCount = 0}. It is not optional. Indeed, the
\tc{Exists} case of \tc{unify} in Figure~\ref{conditional-unification} raises an
error. Liquid Haskell, however, requires functions to be total. We need this
precondition so that Liquid Haskell can prove that this case never occurs.

This totality requirement is not necessary to refinement types in general.
However, in the case of Haskell, laziness lets us write
\begin{verbatim}
{-@ bad :: () -> { false } @-}
bad :: () -> ()
bad _ = let {-@ f :: { false } @-}
            f = error "never happens"
         in (\_ -> ()) f
\end{verbatim}
It may seem that Liquid Haskell could accept this function because \tc{f} appears
to prove \tc{false}. In a strict language this would not be a big problem as
\tc{bad} would loop and any attempt at using \tc{bad} would diverge. But \tc{bad} is
actually a total function. Liquid Haskell rejects \tc{bad} because it fails to
prove that \tc{f} is total, hence refuses to accept its signature.

This is also why the signature of \tc{unify} ends with \tc{/ [formulaSize f]}.
Liquid Haskell needs to prove that \tc{unify} terminates and, because of the
substitutions, \tc{unify} is not a structurally recursive function. So Liquid
Haskell needs a little help in the form of a termination metric. We use here
the number of connectives in the argument formula, which is unaffected by
substitution since we only substitute inside terms.

\subsection{Lemmas in Liquid Haskell}
\label{sec:lh-lemmas}

In the previous sections we have seen that the refined implementation can be
different from the classical version by adding computationally irrelevant
arguments. Another way in which they could differ is with the addition of lemmas.

Take, for instance, the \tc{unifyFormula} function which ties together
\tc{skolemize} and \tc{unify}, it differs from its classical implementation as
follows:

\begin{verbatim}
 unifyFormula :: Set Int -> IntMap (Set Int) -> Formula -> Maybe [(Var, Term)]
 unifyFormula s m f =
     let m' = addSToM s m
-        skf = skolemize s f
+        skf = skolemize s f ? lemmaConsistentSuperset m m' f
         (f'', m'') = runState skf m'
      in unify s m'' f''
\end{verbatim}

This idiom \tc{e?p} means ``use lemma \tc{p} when checking \tc{e}''.
Lemmas are not used automatically, this is how Liquid
Haskell is instructed to use them with parameter values supplied by the user.

Lemmas, in Liquid Haskell, are ordinary functions. Proofs by inductions arise
from ordinary (total!) recursion. In the case of \tc{lemmaConsistentSuperset}
the proof is entirely straightforward

\begin{verbatim}
{-@
lemmaConsistentSuperset
  :: m0:IntMap (SetInt)
  -> {m1:IntMap (Set Int) | intMapIsSubsetOf m0 m1}
  -> {f:Formula | consistentScopes m0 f}
  -> {consistentScopes m1 f}
@-}
lemmaConsistentSuperset
  :: IntMap (Set Int) -> IntMap (Set Int) -> Formula -> ()
lemmaConsistentSuperset m0 m1 (Forall _ f) =
    lemmaConsistentSuperset m0 m1 f
lemmaConsistentSuperset m0 m1 (Exists _ f) =
    lemmaConsistentSuperset m0 m1 f
lemmaConsistentSuperset m0 m1 (Conj f1 f2) =
      lemmaConsistentSuperset m0 m1 f1
    ? lemmaConsistentSuperset m0 m1 f2
lemmaConsistentSuperset m0 m1 (Then (t0, t1) f2) =
      lemmaConsistentSupersetTerm m0 m1 t0
    ? lemmaConsistentSupersetTerm m0 m1 t1
    ? lemmaConsistentSuperset m0 m1 f2
lemmaConsistentSuperset m0 m1 (Eq t0 t1) =
      lemmaConsistentSupersetTerm m0 m1 t0
    ? lemmaConsistentSupersetTerm m0 m1 t1
\end{verbatim}

So straightforward, in fact that the proof was largely written by AI-based code
completion. Since lemmas do not have computational content (\tc{\{p\}} is a
shorthand for \tc{\{\_:() | p \}}), we only care about the existence of a
proof, making code completion particularly useful. Liquid Haskell understanding
the theory of finite maps (see Section~\ref{extending-liquid-haskell}) is crucial in
making this proof so terse.

The lemma \tc{lemmaConsistentSuperset} uses an analogous lemma \tc{lemma\-Consistent\-Superset\-Term} for terms,
whose proof ultimately depends on the following lemma which we must assume of the substitution data type.
Unsurprisingly, the substitution interface needs to satisfy more properties
than in Section~\ref{the-rapier-with-refinement-types} to accommodate
unification variable scopes.

\begin{verbatim}
assume lemmaConsistentSupersetSubst
  :: m0:_
  -> {m1:_ | intMapIsSubsetOf m0 m1}
  -> {s:_ | consistentScopesSubst m0 s}
  -> {consistentScopesSubst m1 s}
\end{verbatim}

\subsection{Extending Liquid Haskell to support \tc{IntMap}}
\label{extending-liquid-haskell}

Our unification case study uses the theory of finite maps. Liquid Haskell,
however does not support a theory of finite maps\footnote{Issue to support maps in
  the Liquid Haskell repository:
  \url{https://github.com/ucsd-progsys/liquidhaskell/issues/2534}}. It is
possible to do without it. In a first approximation we did much of this study in
vanilla Liquid Haskell. But we lost out on automation: we got more lemmas to prove
and pass around. Properties like the scope check, or the lemma
\tc{lemmaConsistentSuperset}, involved operations on finite maps and were more
convoluted.

To support this study, we implemented the theory of finite maps for Liquid
Haskell. It is not ready to integrate in future release yet, for one thing: we only support finite
maps with \tc{Int} as their domain and \tc{Set Int} as their codomain. It could easily be adapted for any fixed
domain and codomain types, but it is not yet a general solution that can be instantiated at any
domain or codomain type. But our ultimate intent is to upstream these changes. Our
modifications can be found in the file \patchfile{ifl25-liquidhaskell.patch} and the file \patchfile{ifl25-liquid-fixpoint.patch}.

The theory of finite maps is a good example of a theory that Liquid Haskell wants
to support: it is both powerful, and widely applicable. Pragmatically, it is also
one that is reasonably easy to support with SMT solvers by translating it to
the theory of arrays.

On the syntax front, Liquid Haskell allows to link a Haskell type with a particular
representation in the SMT solver.

\begin{verbatim}
{-@ embed IntMap * as IntMapSetInt_t @-}
\end{verbatim}

Here we are indicating that \tc{IntMap b} must be represented as \tc{IntMapSetInt\_t}
in the logic. \tc{IntMapSetInt\_t} is an alias for \tc{Array Int (Option (Set Int))}.
An array is an entity that associates keys with values, and which has an equality predicate,
and it is defined as one of the theories in SMT-LIB, the standard interface
to SMT solvers~\cite{BarFT-RR-25}.
The keys in this case are integers, and the values are either \tc{None} if the key
is not in the map, or \tc{Some s} if the key maps to a set \tc{s}. The
\tc{Option} type is a copy of Haskell's \tc{Maybe}.
We do not reuse \tc{Maybe} as Liquid Haskell's framework to connect to the SMT solver is
reused for other languages (\emph{e.g.} \cite{lehmann23}), and we prefer to keep
the implementation free of language specific details.
Here is the declaration of the \tc{Option} data type in SMT-LIB.

\begin{verbatim}
(declare-datatype Option (par (a) (None (Some (someVal a)))))
\end{verbatim}

We arranged for Liquid Haskell to include this declaration in the preamble of any
queries to the SMT solver. The types \tc{Array}, \tc{Int}, and \tc{Set} are already
known to the tooling.
It does not matter what type \tc{b} is instantiated to, the \tc{embed} annotation will
always set the same representation for \tc{IntMap b}, and this is a limitation that
would need to be addressed to support maps properly.

The array theory allows to describe how to retrieve the value associated with
a key, and how to update the value. On the Haskell front, we link these operations
to those of the \tc{IntMap b} type.

\begin{verbatim}
define IntMap.empty = (IntMapSetInt_default None)
define IntMap.insert x y m = IntMapSetInt_store m x (Some y)
define IntMap.lookup x m =
  if (isSome (IntMapSetInt_select m x)) then
    (GHC.Internal.Maybe.Just (someVal (IntMapSetInt_select m x)))
  else
    GHC.Internal.Maybe.Nothing
\end{verbatim}

The operations \tc{IntMapSetInt\_default}, \tc{IntMapSetInt\_store}, and \tc{IntMapSetInt\_select}
are aliases that we implemented in Liquid Haskell to call to the array operations.
In the case of \tc{lookup}, we translate the \tc{Option} type to Haskell's \tc{Maybe}.

The implementation of union, intersection,
difference, and subset checks for maps, however,
need operations beyond the standard interface, and not all SMT solvers can support
them. In our implementation we used the \tc{map} operation of the
Z3 SMT solver. The following snippet contains the implementation of
\tc{intMapIsSubsetOf} in SMT-LIB, and we also feed these declarations to the
SMT solver in a preamble to the queries.

\begin{verbatim}
; Similar to do {a0 <- oa0; a1 <- oa1; guard (a0 /= a1); pure a0}
(define-fun difference_strict_p2p
  ((oa0 (Option (Set Int)))
   (oa1 (Option (Set Int))))
  (Option (Set Int))
  (match oa0
    ((None None)
     ((Some a0) (match oa1
                  ((None oa0)
                   ((Some a1) (ite (= a0 a1) None oa0))))))))

; Similar to: empty == zipWith difference_strict_p2p xs ys
; where zipWith applies the function pointwise to the values in the
; arrays
(define-fun IntMapSetInt_isSubsetOf
  ((xs (Array Int (Option (Set Int))))
   (ys (Array Int (Option (Set Int)))))
  Bool
  (= ((as const (Array Int (Option (Set Int)))) None)
     ((_ map IntMapSetInt_difference_strict_p2p) xs ys)))
\end{verbatim}

Besides the limitation of the \tc{embed} annotation, another barrier for
proper support is that old versions of SMT-LIB require user defined
functions to have monomorphic types. This means, for instance, that
the type of \tc{IntMapSetInt\_isSubsetOf} cannot be generalized to work
on any \tc{IntMap}.

While newer versions of the standard allow
for polymorphic types, these still need to be implemented by SMT solvers.
Until the implementations catch up with the standard, feeding operations with
monomorphic types will require Liquid Haskell to be smart about generating
these operations with the appropriate types, instead of putting them in a
preamble once and for all queries.

\section{Evaluation}
\label{evaluation}

The substitution case study of Section~\ref{capture-avoiding-substitution}
allows for a direct comparison between type methods and refinement type methods. We
can see that the trusted code base of the Liquid Haskell version of
Section~\ref{the-rapier-with-refinement-types} is quite small compared to that
of the foil~\cite{maclaurin23} (reviewed in
Section~\ref{the-rapier-with-stronger-types}). This is in large part
because refinement types can enforce invariants without the need for
abstract types, and such an open interface can be extended by the user.
Contrast with the abstract-type approach where you have to design, upfront, a
set of invariant-preserving operations sufficient to express downstream programs.
None of these functions will benefit from the abstract types invariant, hence
will be part of the trusted code base. Even when we mix refinement and abstract
types as in Section~\ref{ensuring-the-scope-set-is-checked}, we do not have quite
as large a trusted code base to consider.

This is not to mean that refinement types are superior to type abstractions.
They are best at enforcing different types of invariants, as discussed in
Section~\ref{ensuring-the-scope-set-is-checked}.

When the invariants of a program naturally involve mathematical objects such as
arithmetic or sets, refinement types are likely to be more approachable,
requiring less careful a design than coming up with an encoding inside and
ML-like type system. Proposing refinement type signatures requires
determining appropriate invariants for a task, which is a requisite for any static
checking approach. But it doesn't impose the burden of encoding the invariants with lower-level
constructs. On the other hand, when a program needs a theory that Liquid
Haskell, say, does not have support for, it may not be that clear and the program
author may need to mobilize comparable effort for refinement types as she would
have for an abstract-type encoding.

\paragraph{Error reporting} A type-checker approach, however, is likely to produce error messages that
are easier both to understand and to fix, provided that the user goal is feasible.
The user is guided into correcting the errors
by the types and the operations of the supporting library. With SMT solvers,
there is always the question of whether a goal is provable or not in the
theories at hand. Is there some additional lemma that is necessary about the user defined
functions? The user has to figure it out on her own. How are the assumptions
insufficient to prove the goal? The user has to compute it on her own too,
although it is plausible that counterexamples or better location information~\cite{webbers24}
can be offered when the tooling matures.

But there are informative error messages too. Let us consider the lemma \tc{lemma\-Consistent\-Scopes\-Subst} discussed in
Section~\ref{checking-unifyEq}.
If we drop this lemma from the definition of \tc{unifyEq}, we get the following
error message (heavily edited for presentation):
\begin{verbatim}
publications/ifl25-rtdd/examples/Unif2.hs:580:18: error:
    Liquid Type Mismatch
    The inferred type
      ss' : {ss' : Subst {v : Term | consistentScopesTerm m v} |
                 Set_com Set.empty == domain ss'}
    is not a subtype of the required type
      VV : {VV : Subst Term | consistentScopesSubst m VV}
    in the context
      ?g : {?g : Maybe (Subst Term) |
               ?g == Just ss'
            && ?g == inverseSubst s m
                       (narrowForInvertibility (freeVars t1) ss)}

      t0 : {t0 : (Int, (Subst Term)) | t0 == SA (i, ss)
               isSubsetOf (freeVars t0) s
            && consistentScopesTerm m t0}

      t1 : {t1 : Term |
               isSubsetOf (freeVars t1) s
            && consistentScopesTerm m t1}

      i : Int
      s : Set Int
      m : IntMap (Set Int)
      ss : Subst Term
    Constraint id 168
    |
578 |     , let t' = substitute (freeVarsSubst ss') m ss' t1
    |                                                 ^^^
\end{verbatim}

We can get quickly that the predicate in the required type is one of the
conjuncts in the refinement type of
a parameter of \tc{substitute}. That is \tc{ConsistentScopedSubst}, a type alias
we declared in the same module, and in this case expands as follows.

\begin{verbatim}
     {ss':Subst Term |
           isSubsetOf (freeVarsSubst ss') (freeVarsSubst ss')
        && consistentScopesSubst m ss'
     }
\end{verbatim}

To get at the missing lemma, in this case we only need to connect the predicates
in the inferred and the required refinement types. Let us prune the irrelevant
bits from the error message first.

\begin{verbatim}
    The inferred type
      ss' : Subst {v : Term | consistentScopesTerm m v}
    is not a subtype of the required type
      VV : {VV : Subst Term | consistentScopesSubst m VV}
\end{verbatim}

And then we can substitute \tc{VV} by \tc{ss'} in the goal, which gives
pretty much the lemma statement.

\begin{verbatim}
    The inferred type
      ss' : Subst {v : Term | consistentScopesTerm m v}
    is not a subtype of the required type
      ss' : {ss' : Subst Term | consistentScopesSubst m ss'}
\end{verbatim}

When there are static check failures, insight is often necessary to identify
a missing lemma or a missing precondition.
Recursive functions like \tc{skolemize} start with a core
set of conjuncts that sometimes needs to be grown as static checks reveal the need of
stronger postconditions for the result of the recursive calls.

\paragraph{Maturity} Maybe relatedly, the maturity of Liquid
Haskell is rather lacking still. We have encountered a
non-negligible number of bugs (18) in the Liquid Haskell tooling and usability issues
while conducting our study. Our source code contains comments explaining the defects
where we were affected.
The sources of most of these defects seem to locate in the Liquid Haskell
implementation rather than the SMT solver, and there was an issue encountered in the SMT
solver\footnote{We found a problem in the Z3 SMT solver, which sprung some follow up issues
further linked in the original issue: \url{https://github.com/Z3Prover/z3/issues/7770}}.
Fortunately, none of them look very
difficult to address, but they do have a severe impact on user experience in aggregate.

Besides, Liquid Haskell lacks support for many standard features of Haskell.
In our code we have been using the simplest possible style of programming.
There are no GADTs, no type families, and minimal use of type classes (since
Liquid Haskell has some support for type classes~\cite{liu20}). At the moment,
pushing for more demanding programming patterns is likely to surface more
inconveniences. Aiming for the simplest style is, therefore, a pragmatic constraint of
the current implementation. For further insight on the challenges of using Liquid Haskell,
Gamboa et al.~\cite{gamboa25} report on a study that collects the voices
of its users.

On the performance front, all of the SMT-LIB queries in the unification example run
in 11 seconds, 0.04 seconds for \tc{Subst2.hs}, and 0.03 seconds in \tc{Subst1.hs}.
That is sometimes faster than compiling a module with the GHC compiler.
Where things get slower is when measuring Liquid
Haskell end-to-end, which spends several seconds checking the examples and interacting with the
SMT solver (3 minutes when checking unification, 4 seconds checking \tc{Subst2.hs},
1.5 seconds checking \tc{Subst1.hs}). The authors deem that performance of Liquid Haskell
can be improved to approach that of the SMT solver queries, and probably further by
reducing the number of queries.

\paragraph{Composability} Perhaps one of the biggest compromises when encoding properties in the type-checker
is that one needs to narrow the expressible properties to a feasible set that allows
to write a supporting library. If we wanted to have static checks like those of the
unification example, we would need new type encodings. Or in other words, new type indices
need to be conceived to relate the parameters of our functions.

$$\begin{array}{l@{\ }l@{\ }l}
    \mathit{skolemize} & \mathit{::}  & \mathit{Scope}\ s_1 \ldots s_n \\
                       & \rightarrow  & \mathit{Formula}\ f_1 \ldots f_j \\
                       & \rightarrow  & \mathit{State}\ t_1 \ldots t_k\ (\mathit{Scope}\ e_1 \ldots e_l)\ (\mathit{Formula}\ o_1 \ldots o_m)
\end{array}$$

Then there would be the effort of writing a library, and later on there would be the
effort of composing the encodings of different libraries when more than one such
is needed. Suppose we started with the static checks to avoid name captures as in
Section~\ref{capture-avoiding-substitution}, and we wanted to add the scopes checks
required to deal with unification variables. With refinement types we need to add
the corresponding conjuncts to the refinement types, and perhaps some \textit{phantom}
parameter like \tc{m} here.

\begin{verbatim}
substituteFormula
  :: s:Set Int
  -> m:IntMap (Set Int)
  -> ss:ConsistentScopedSubst s m
  -> {f:ScopedFormula (domain ss) | consistentScopes m f}
  -> {v:ScopedFormula s |
          formulaSize f == formulaSize v
       && consistentScopes m v
       && existsCount v = existsCount f
     }
\end{verbatim}
Besides the usual scope checks, we are checking that the size of the formula
is preserved, that the amount of existential binders is preserved, and that
the unification scopes in the output are those in the input formula and in
the range of the substitution. We also check that substitution preserves
the consistency of the unification scopes.

\section{Comparable systems}

Liquid Haskell is not the only tool reaching to SMT solvers for static checks.
The most similar tool is F*~\cite{swamy16}, which is based on a refinement type
system as well. Another family of related systems are those with Hoare-style
pre- and post-conditions to functions such as Why3~\cite{filli13} and
Dafny~\cite{leino17} (impure functional programming languages), or
ESC/Java~\cite{flanagan02} and Frama-C~\cite{kosmatov24} (imperative languages).

All of the above systems could have served as a vehicle for our case study,
though the further we go down that list, the more different the language is too
Liquid Haskell, and the more adaptation that would require. The type systems
also get weaker and the latest the language is in the list, the more one has to
lean on the SMT for static checks.

\section{Conclusions}
\label{conclusions}

The tooling is not ready for widespread use. Yet it is plausible that in a decently
close future, we have access to SMT solvers and refinement-types to assist us in
our programming.

Refinement types enable a more direct expression of properties,
particularly when the SMT solver supports the relevant theories. Reasoning
mechanisms are reused from the existing tooling, instead of encoding them
in the type checker. This makes easier both to enforce our own invariants and to
compose properties coming from different sources.

The generality of the approach, and the simplicity with which it enables
composition of different properties, are unique features that make it a strong
candidate to impact programming practice in the future.

Through our two case studies, we have tried to make a first step in
understanding how we will be best able to leverage future such tools, even
in situations where we can manage to use current type-checkers today. As a closing
note, let us reproduce the principles that we have proposed throughout the article.

\vspace{-0.8cm}
\listofkeytheorems[
  ignoreall,
  show={principle},
  print-body,
  title={},
  ]

\bibliographystyle{ACM-Reference-Format}
\bibliography{references}

\end{document}